\documentclass[twocolumn]{jpsj2} 
\title{Spin Susceptibility, Phase Diagram, and Quantum Criticality in the Eelctron-Doped High $T_c$ Superconductor Ba(Fe$_{1-x}$Co$_{x}$)$_{2}$As$_{2}$}

\author{Fanlong \textsc{Ning}$^{1}$, Kanagasingham \textsc{Ahilan}$^{1}$, Takashi  \textsc{Imai}$^{1,2}$\thanks{E-mail: imai@mcmaster.ca}, Athena S.  \textsc{Sefat}$^{3}$, Ronying  \textsc{Jin}$^{3}$, Michael A.  \textsc{McGuire}$^{3}$, Brian C.  \textsc{Sales}$^{3}$, and David  \textsc{Mandrus}$^{3}$}

\inst{$^{1}$Department of Physics and Astronomy, McMaster University, Hamilton, Ontario L8S4M1, Canada\\
$^{2}$Canadian Institute for Advanced Research, Toronto, Ontario M5G1Z8, Canada \\
$^{3}$Materials Science and Technology Division, Oak Ridge National Laboratory, TN 37831, USA}

\abst{We report a systematic investigation of Ba(Fe$_{1-x}$Co$_{x}$)$_{2}$As$_{2}$ based on transport and $^{75}$As NMR  measurements, and establish the electronic phase diagram.  We demonstrate that doping progressively suppresses the uniform spin susceptibility and low frequency spin fluctuations.  The optimum superconducting phase emerges at $x_{c}\simeq0.08$  when the tendency toward spin ordering completely diminishes.  Our findings point toward the presence of a quantum critical point near $x_{c}$ between the SDW (spin density wave) and superconducting phases.}

\kword{iron pnictide superconductor, high temperature superconductivity, NMR, quantum criticality}

\begin{document}
\maketitle

The recent discovery of iron-pnictide high $T_c$ superconductors \cite{kamihara,sefat1,sm,ren,eisaki,rotter} poses a new intellectual challenge in condensed matter physics.  Although the superconducting mechanism remains unknown, it has become apparent that iron-pnictides share remarkable similarities with high $T_c$ cuprates: the undoped parent phases LaFeAsO and BaFe$_2$As$_2$  have FeAs square-lattice sheets, and itinerant electrons in these layers undergo antiferromagnetic long range order (AFLRO) below modest temperatures, $T_{SDW} \sim 140$~K\cite{nakai, delacruz,huang}; electron or hole doping suppresses the AFLRO and induces superconductvity.  There exist clear dissimilarities, too.  For example, all five 3d orbitals of Fe atoms contribute to the multiple Fermi surfaces, hence to superconductivity, in iron-pnictides\cite{singh,ding}.  In contrast, only the Cu 3d$_{x^{2}-y^{2}}$ orbital plays a role in curates.  Furthermore, substitution of Co atoms into Fe sites of the parent phases results in electron doping and induces superconductivity \cite{sefat2,sefat3,jasper} without creating localized moments,\cite{ning} while Zn$^{2+}$ ions doped into Cu$^{2+}$ sites induce localized moments and destroy superconductivity in cuprates.   Sorting out these similarities and dissimilarities may lead us to an understanding the mechanism of high $T_c$ superconductivity in iron-pnictides as well as in cuprates.  

In this {\it Letter}, we utilize transport and NMR techniques to probe the evolution of the electronic properties of Ba(Fe$_{1-x}$Co$_{x}$)$_{2}$As$_{2}$ single crystals from the undoped SDW phase ($x=0$ with $T_{SDW}=135$~K), underdoped SDW phase ($x=0.02$ with $T_{SDW}=100$~K, and $x=0.04$ with $T_{SDW}=66$~K), optimally doped superconducting phase ($x=0.08$ with $T_{c}=22$~K),  to the slightly overdoped regime ($x=0.105$ with $T_{c}=15$~K).  We establish the electronic phase diagram which has a quantum critical point at $x_{c} \sim 0.08$, and explore the possible relation between paramagnetic spin fluctuations and the mechanism of superconductivity.  From NMR Knight shift and spin-lattice relaxation rate measurements, we show that electron doping progressively suppresses the uniform spin susceptibility $\chi_{spin}$ and low frequency spin fluctuations.  The optimally doped superconducting phase emerges at $x=0.08~(\sim x_{c})$ as soon as doped electrons completely suppress the tendency toward AFLRO.  

We grew single crystals of Ba(Fe$_{1-x}$Co$_{x}$)$_{2}$As$_{2}$ from FeAs flux,\cite{sefat3} and determined the actual Co concentration $x$ by EDS (Energy Dispersive X-ray Spectroscopy).  The $x=0.08$ sample is the same specimen as the one we previously studied and labeled as $x=0.1$,\cite{sefat3,ning,ahilan} and has the nominal starting composition of $x=0.1$.   The bulk characterization of the $x=0.04$ and 0.08 crystals is discussed elsewhere \cite{sefat3,ahilan}.  We carried out NMR measurements using standard pulsed NMR techniques for aligned single crystals with the total mass of $\sim2$ to $\sim20$~mg.  We measured the frequency dependence of the spin-lattice relaxation rate divided by temperature $T$, $\frac{1}{T_{1}T}$, within a single NMR line by integrating the FFT signals for an appropriate frequency range.   AC resistivity measurements were performed with the standard four probe method.  We define the critical temperature $T_{SDW}$ of AFLRO based on the maximum negative slope of $\rho_{ab}$,\cite{ahilan} while $T_c$ was determined by SQUID measurements.  We summarize the resistivity data in Fig.1, and the electronic phase diagram of $T_{SDW}$ and $T_c$ in Fig.2.

In Fig.3, we present typical $^{75}$As NMR lineshapes of the $I_{z}=-1/2$ to $+1/2$ transition  at 175~K obtained by FFT of the spin echo envelope measured in a fixed magetic field of $B_{ext}\sim 7.7$~Tesla applied along the crystal c-axis.  For clarity, we converted the horizontal axis of the NMR resonance frequency $f$ to the NMR Knight shift, $K$, based on the following relation;\cite{nuq}
\begin{equation}
f  = (1+K)\gamma_{n} B_{ext},
\end{equation}
where $\gamma_{n}$ is the nuclear gyromagnetic ratio, and
\begin{subequations}
\begin{align}
K & =K_{spin}+K_{chem},\\
K_{spin} & = \frac{A_{hf}}{N_{A}\mu_{B}}\chi_{spin}.\
\end{align}
\end{subequations}
The Knight shift $K$ consists of two separate contributions.  The first term in Eq.2a, $K_{spin}$, is the spin contribution, and proportional to the local spin susceptibility along the c-axis, $\chi_{spin}$, as defined in Eq.2b.  The second term in Eq.2a, $K_{chem}$, is a temperature-independent background term ({\it the chemical shift}), and is not related to $\chi_{spin}$.  From the measurement at 4.2~K in the superconducting state below $T_c$ of the $x=0.08$ sample, we estimate $K_{chem} = 0.224$~\% along the c-axis.\cite{ning}     $A_{hf} = 18.8$~kOe/$\mu_{B}$ in Eq.2b represents the hyperfine interaction along the c-axis between the observed $^{75}$As nuclear spin and electron spins,\cite{takigawa} and $N_{A}$ is the Avogadro's number.

We summarize the $T$ and $x$ dependences of the center of gravity of $K$ in Fig.4.  Eq.2b implies that a change of the spin contribution to the Knight shift by the amount of $\Delta K_{spin} = 0.0339$~\% corresponds to that of $\Delta \chi_{spin} = 1 \times 10^{-4}$~emu/mol-Fe along the c-axis.  We show the conversion on the right vertical axis of Fig.4.  The magnitude of $\chi_{spin}$ is comparable to that of  LaFeAsO$_{0.9}$F$_{0.1}$.\cite{imaiTokyo}  Notice that the origin on the right axis, $\chi_{spin} = 0$,  is matched with the temperature independent background contribution $K_{chem} = 0.224$\%.  Fig.4 represents the first successful measurements of the intrinsic behavior of $\chi_{spin}$ for a broad range of concentrations of iron-pnictides.

We notice two striking features in Fig.4.  First, doped electrons {\it suppress} $\chi_{spin}$.  We recall  that  doped holes {\it enhance}  $\chi_{spin}$ in high $T_c$ cuprates,\cite{johnston}  and the enhancement is interpreted as a consequence of the suppression of some energy scale with doping, such as the effective spin-spin exchange interaction $J$ or spin pseudo-gap $\Delta_{PG}$.\cite{johnston,timusk}  If we interpret the results of Fig.4 in an analogous manner, our results imply that the energy scale {\it increases} with doping in  Ba(Fe$_{1-x}$Co$_{x}$)$_{2}$As$_{2}$.  Second, the $T$ dependence of $\chi_{spin}$ is qualitatively similar in the paramagenetic state of all samples from undoped antiferromagnet ($x=0$) to the slightly overdoped superconductor ($x=0.105$);  $\chi_{spin}$ decreases roughly linearly for all samples below 300~K, then begins to level off  below $\sim 100$~K toward $T_c$ in the superconducting samples with $x \geqslant 0.08$.   $\rho_{ab}$ also shows a T-linear behavior in the same temperature regime below $\sim 100$~K, as initially reported for $x=0.08$.\cite{sefat3,ahilan}  That is, {\it the high $T_c$ superconductivity of iron-pnictides arises from a novel electronic state with $\chi_{spin} \sim constant$ and $\rho_{ab}\sim T$ after $\chi_{spin}$ is suppressed.}

What are the dominant mechanisms that govern the $T$ and $x$ dependences of $\chi_{spin}$?  Since the undoped sample undergoes AFLRO below $T_{SDW}=135$~K, one obvious possibility is that the growth of antiferromagnetic short-range order (AFSRO) suppresses $\chi_{spin}$ with decreasing $T$.   However,  AFSRO alone can't account for the $x$ dependence, because $\chi_{spin}$ is smaller for the Co doped samples without AFLRO.  In other words, if we attribute $T$ and $x$ dependencies entirely to the effects of AFSRO, we would be led to an unphysical conclusion that the antiferromagnetic spin-spin correlation is stronger in the Co doped superconducting phase than in the undoped phase with AFLRO.  Moreover, growth of AFSRO generally results in enhancement of $1/T_{1}T$ with decreasing $T$, as observed for La$_{2-x}$Sr$_{x}$CuO$_{4}$,\cite{imai} but our $1/T_{1}T$ data contradict with such a scenario as explained below.  While AFSRO is very likely to play a significant role in controlling the behavior of $\chi_{spin}$, there must be an additional mechanism which suppresses  $\chi_{spin}$.  We will address this issue below.

Before proceeding, we would like to comment briefly on the implications of the systematic line broadening of the NMR lines with doping  in Fig.3.\cite{nuq}   Eq.2b implies that the distribution of $K_{spin}$ in Fig.3 reflects that of $\chi_{spin}$.  Accordingly, the FFT lineshapes in Fig.3 represent a histogram of the distribution of  $\chi_{spin}$ in each sample.
For example, some parts of the $x=0.08$ sample have $\chi_{spin}$ as large (small) as that of the $x=0.04$ ($x=0.105$) sample, i.e. the local electronic properties are inherently inhomogeneous at a microscopic level.   We emphasize that one can't attribute the present finding to macroscopic inhomogeneity of the Co concentrations.  As shown in Fig.1,  the electrical resistivity $\rho_{ab}$ is very sensitive to phase transitions at $T_{SDW}$ and $T_c$.  If the $x=0.08$ crystal had a macroscopic domain with Co 4\% doping, for example,  we would observe an additional SDW anomaly in the $\rho_{ab}$ data.   We recall that analogous microscopic inhomogeneity was also observed for the hole-doped high $T_c$ cuprates La$_{2-x}$Sr$_{x}$CuO$_4$,\cite{singer1,singer2} and understood as the consequence of microscopic patch-by-patch inhomogeneity of the carrier concentration with the typical length scales of $\sim 3$~nm.  Our results in Fig.3 show that analogous "patchy" electronic inhomogeneity also exists in Ba(Fe$_{1-x}$Co$_{x}$)$_{2}$As$_{2}$ with comparable length scales.  

Additional clues on the nature of spin correlations may be seen in  the $T$ and $x$ dependencies of $1/T_{1}T$.  We summarize $1/T_{1}T$ measured at the peak of the NMR lineshapes in Fig.5.    Theoretically,  $1/T_{1}T$ is related to the wave vector integral of the low frequency component of spin fluctuations,
\begin{equation}
\frac{1}{T_{1}T} \sim  \sum_{\bf q} |\gamma_{n}A({\bf q})|^{2} \frac{Im~\chi({\bf q},f)}{f},
\end{equation}
where $Im~\chi({\bf q},f)$ is the imaginary part of the dynamic electron spin susceptibility at the NMR frequency $f$ ($\sim 65$~MHz), $|A({\bf q})|^{2} =  |A cos(q_{x}a/2)cos(q_{y}a/2)|^{2}$ is the form factor of the transferred hyperfine interaction at the $^{75}$As sites ($a$ is the lattice constant),\cite{ning} and the wave vector summation of ${\bf q}$ is taken over the first Brillouirn zone.  The undoped $x=0$ sample does not exhibit a large enhancement of $1/T_{1}T$ immediately above $T_{SDW}=135$~K, in agreement with Kitagawa et al.\cite{takigawa}.  There are two reasons for the absence of the signature of critical slowing down.  First, the AFLRO is accompanied by a tetragonal to orthorhombic structural phase transition,\cite{huang} and the simultaneous magnetic and structural phase transitions are weakly first order.\cite{takigawa}   Second, since we are applying $B_{ext}$ along the c-axis, the ab-plane components of the dipole hyperfine fields from Fe moments, as well as the transferred hyperfine fields in Eq.3, cancel out at the location of $^{75}$As sites when spin correlations are commensurate\cite{delacruz,huang} with the lattice.

On the other hand, underdoped $x=0.02$ and 0.04 samples show a divergent behavior of $1/T_{1}T$ due to critical slowing down of spin fluctuations toward $T_{SDW}=100$~K and 66~K, respectively.  We also conformed that the paramagnetic NMR signals disappear below $T_{SDW}$.  These results strongly suggest the SDW state, presumably incommensurate with the lattice, emerges below $T_{SDW}$ as a result of the second order phase transition.

We have not conducted NMR measurements in the concentration range $x=0.05 \sim 0.07$.  However, the NMR lineshapes of the $x=0.04$ and 0.08 samples are significantly superposed because of the patchy inhomogeneity, and we have found that $1/T_{1}T$ measured at the half-intensity position on the lower $K$ side of the $x=0.04$ is nearly the same as  $1/T_{1}T$ measured at the half intensity position on the higher $K$ side of the $x=0.08$ crystals, see Fig.3.  These  $1/T_{1}T$ data points measured in the middle between the peaks of $x=0.04$ and 0.08 are plotted in Fig.5, labeled as the results for the effective concentration of  "6\%."   The enhancement of $1/T_{1}T$ toward $T_{c} = 22$~K for "6\%" and 8\% samples signals the residual effects of low frequency AFSRO at low temperatures, and we conclude that the FeAs layers are on the verge of  forming an SDW ground state even for $x\lesssim0.08$. 

In contrast, $1/T_{1}T$ for the slightly overdoped $x= 0.105$ levels off after monotonic decrease with $T$, and exhibits no evidence for the enhancement due to low frequency AFSRO.  Equally important to notice is that $1/T_{1}T$ initially decreases with $T$ below 300~K even in the underdoped $x=0.02$ and 0.04 samples.  We confirmed that $1/T_{1}T$ at the Co sites in $x=0.08$\cite{ning} and $x=0.04$ (not shown) shows qualitatively the same behavior as at $^{75}$As sites; hence the present results are not the consequence of accidental cancellation of the hyperfine fields at  $^{75}$As sites.  Recalling that  the uniform (i.e. ${\bf q}={\bf 0}$) spin susceptibility $\chi_{spin}$ is also suppressed below 300~K for all concentrations, we conclude that the low energy spin excitations are suppressed with decreasing $T$ except near $T_{SDW}$, i.e.  Ba(Fe$_{1-x}$Co$_{x}$)$_{2}$As$_{2}$ exhibits {\it spin pseudo-gap} behavior for a broad concentration range.  We note that analogous pseudo-gap behavior was first observed in LaFeAsO$_{0.89}$F$_{0.11}$ ($T_{c}=28$~K) using $^{19}$F NMR\cite{ahilanNMR}.  Although the mechanism behind the spin pseudo-gap behavior is unclear at this time, it is interesting to realize that the $J_{1}-J_{2}$ Heisenberg model, which is considered a viable starting point for the theoretical description of iron-pnictides,\cite{si} is indeed expected to exhibit a gapped behavior in NMR properties.\cite{mila}  On the other hand, if we consider Fe spins as localized moments described by the $J_{1}-J_{2}$ Heisenberg model,  we expect that the exchange narrowing effects lead to $1/T_{1} \sim constant$,\cite{imai} hence $1/T_{1}T \sim 1/T$, in the intermediate temperature range $T \sim J_{1,2}/2$ or higher.  No evidence for such  localized moment behavior is seen in Fig.5 at least up to 300~K.

How does electron doping affect the spin pseudo-gap?  In Fig.4, we fit the temperature dependence of $K$ by assuming an empirical activation formula, $K = A+B \times exp(-\Delta_{PG}/k_{B}T)$, where $A$ and $B$ are constants.  The best fit yields $\Delta_{PG}/k_{B} \simeq 710$~K, 570~K, 520~K, and 490~K for $x=0$, 0.04, 0.08, and 0.105.  It is not clear whether the apparent systematic decrease of $\Delta_{PG}$ with doping is intrinsic, because we found that the fit is extremely sensitive to the choice of the constant term $A$ and the uncertainty of $\Delta_{PG}$ is as large as $\sim30$~\%.   For LaFeAsO$_{1-x}$F$_{x}$, analogous fit of $1/T_{1}T$ or $K$ yields $\Delta_{PG}/k_{B} \sim 160$~K for $x\geqslant 0.10$.\cite{nakai,nakai2,imaiTokyo}

In order to gain insights into the essential physics of the suppression of low energy spin excitations without making any assumptions, we create the contour map of equal values of $K$ or $1/T_{1}T$ in Fig.2.  First, let us focus our attention on the black dotted lines representing the equal value contours of $K$.  We can clearly observe, without relying on any theoretical models or assumptions, that $\chi_{spin}$ is suppressed for lower $T$ and larger $x$.  Turning our attention to the blue dashed lines representing the equal value contours of $1/T_{1}T$, we also observe that the residual tendency toward forming the SDW ground state barely disappears at the optimum doping $x=0.08$ with the highest $T_{c}=22$~K.  Thus we conclude that the "superconducting dome" appears right on top of the quantum critical point $x_{c} \sim 0.08$ where the tendency toward forming the SDW phase disappears.  We note that the disappearance of the low frequency components of spin fluctuations does {\it not} mean that spin fluctuations diminish all together.  In fact, the pseudo-gap behavior suggests that the over all spectral weight of spin fluctuations may be shifted to higher energies.   

T.I acknowledges financial support by NSERC, CFI and CIFAR.  Work at ORNL was supported by the Division of Materials and Engineering, Office of Basic Sciences.  A portion of this work was performed by Eugene P. Wigner Fellows at ORNL.\\

\section*{Figure Captions}

Fig.1\\
The in-plane resistivity $\rho_{ab}$.  Vertical solid arrows mark $T_{SDW}$.\\

Fig.2\\
The electronic phase diagram of Ba(Fe$_{1-x}$Co$_{x}$)$_{2}$As$_{2}$.  The filled circles and triangles represent $T_{SDW}$ and $T_c$, respectively.  The blue dashed lines and black dotted lines represent the equal magnitude contours of $1/T_{1}T$ and the Knight shift $K$, respectively. \\

Fig.3\\
Representative $^{75}$As NMR lineshapes obtained by FFT at 175~K.  We converted the horizontal frequency axis to that of $K$ using Eq.2.    Also plotted is $1/T_{1}T$ measured at various points within each FFT line.  The horizontal bars specify the range of $K$ in which the FFT intensity was integrated for $1/T_{1}T$ measurements.  \\

Fig.4\\
The center of gravity of the $^{75}$As NMR Knight shift $K$ measured along the crystal c-axis.  $K=0.224$~\% corresponds to the background arising from the chemical shift $K_{chem}$.  The conversion to spin susceptibility $\chi_{spin}$ based on Eq.2b is shown on the right vertical axis.  Vertical solid and dashed arrows mark $T_{SDW}$ and $T_c$, respectively.  Solid curves are a fit to an activation formula with $\Delta_{PG}/k_{B}= 711$~K ($x=0$), 570~K ($x=0.04$), 520~K ($x=0.08$), and 490~K ($x=0.105$).\\

Fig.5\\
Filled symbols show $1/T_{1}T$ measured at the peak of the $^{75}$As NMR lineshapes for 0, 2, 4, 8, and 10.5\% doped samples.  Orange open circles (triangles) labeled as $"6\%"$ were deduced from the half intensity position of the FFT lineshape of the 8\% (4\%) doped sample on the higher (lower) frequency side.  See the main text for details.


\begin{thebibliography}{99}
\bibitem{nuq} The second order qudrupole effects are negligibly small, and the results in Fig.3 do not depend on the magnitude of $B_{ext}$.
\end{thebibliography}
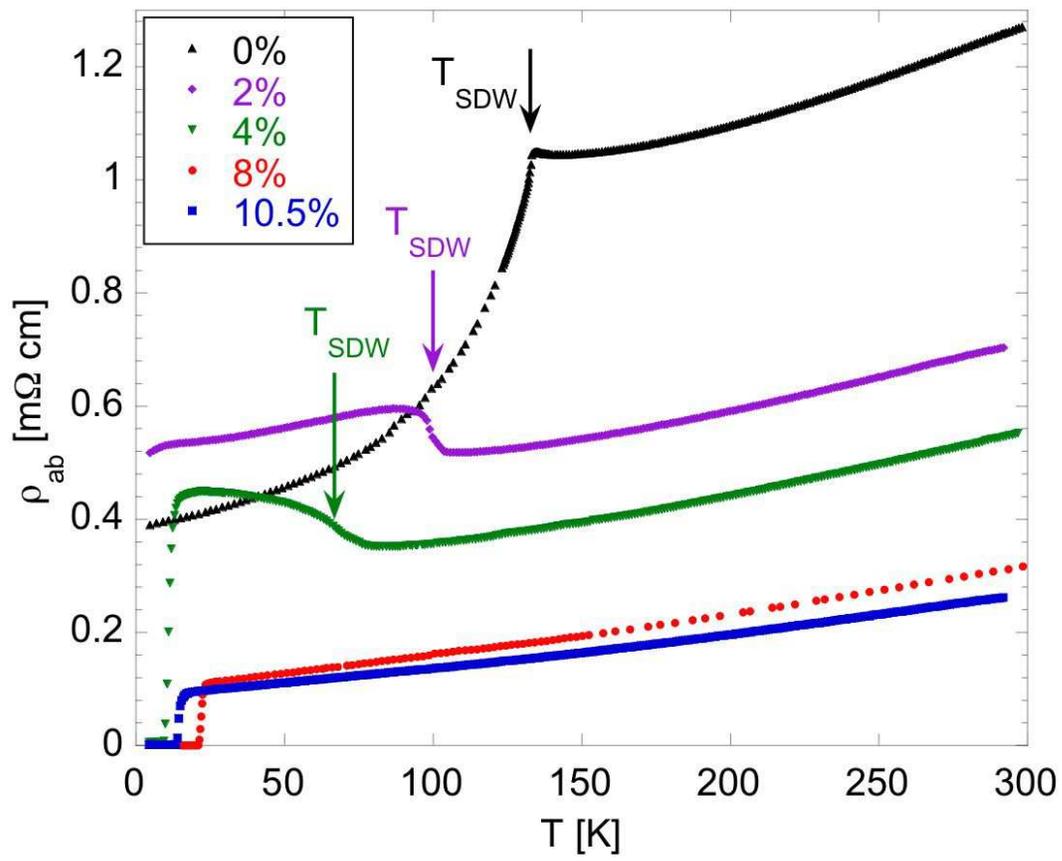
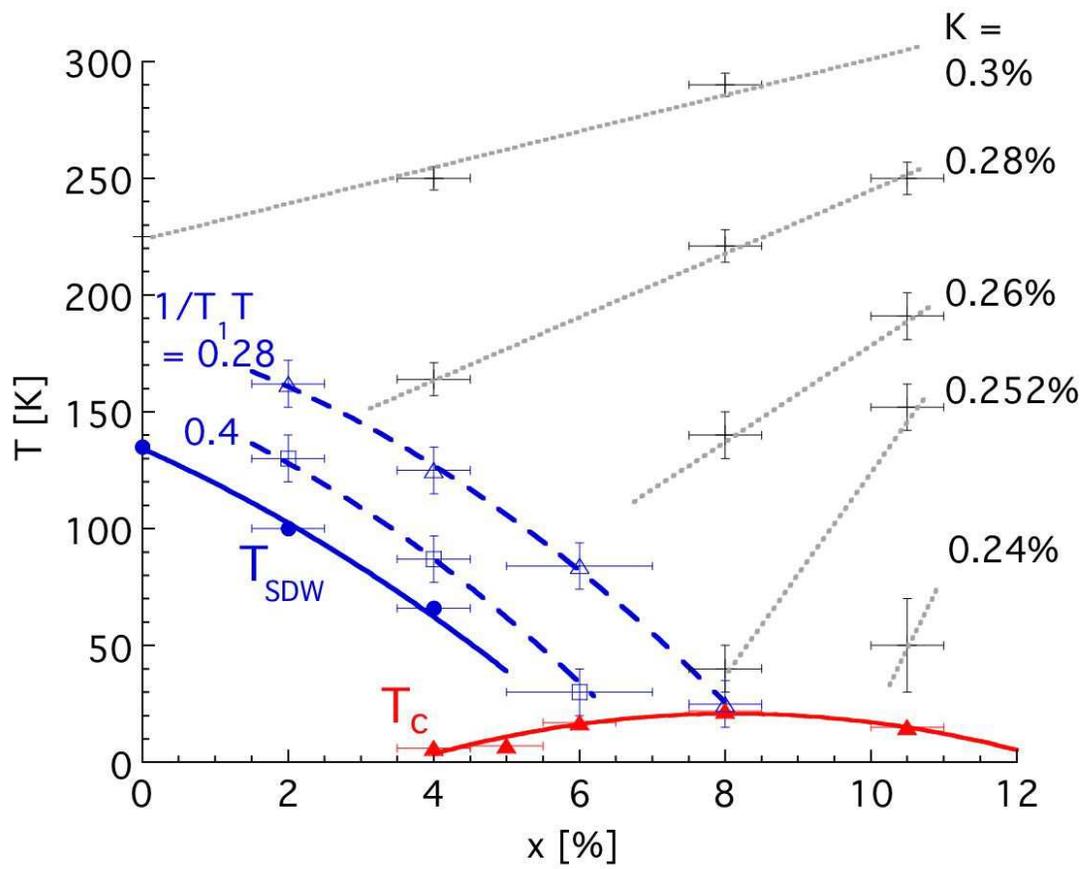
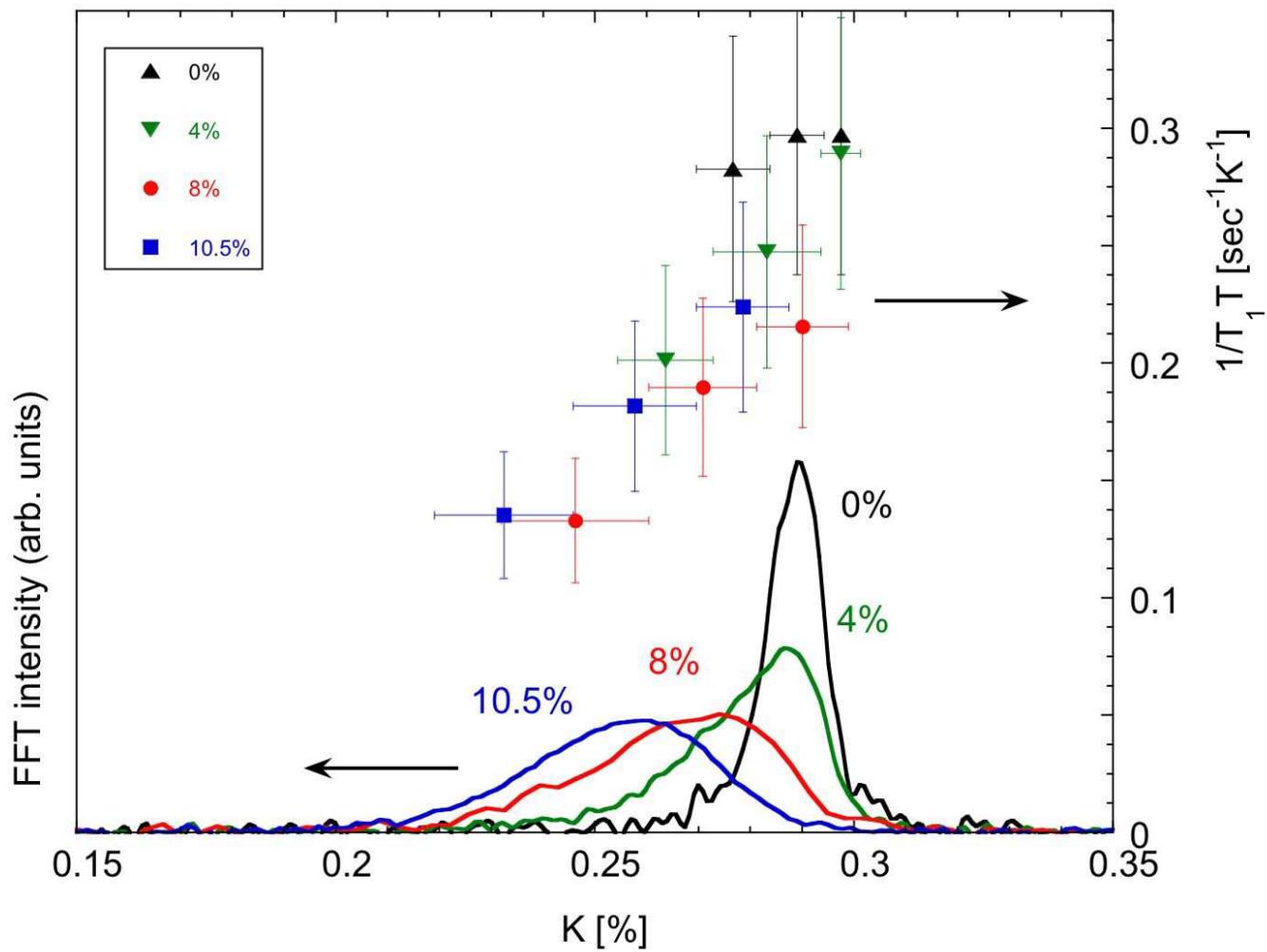
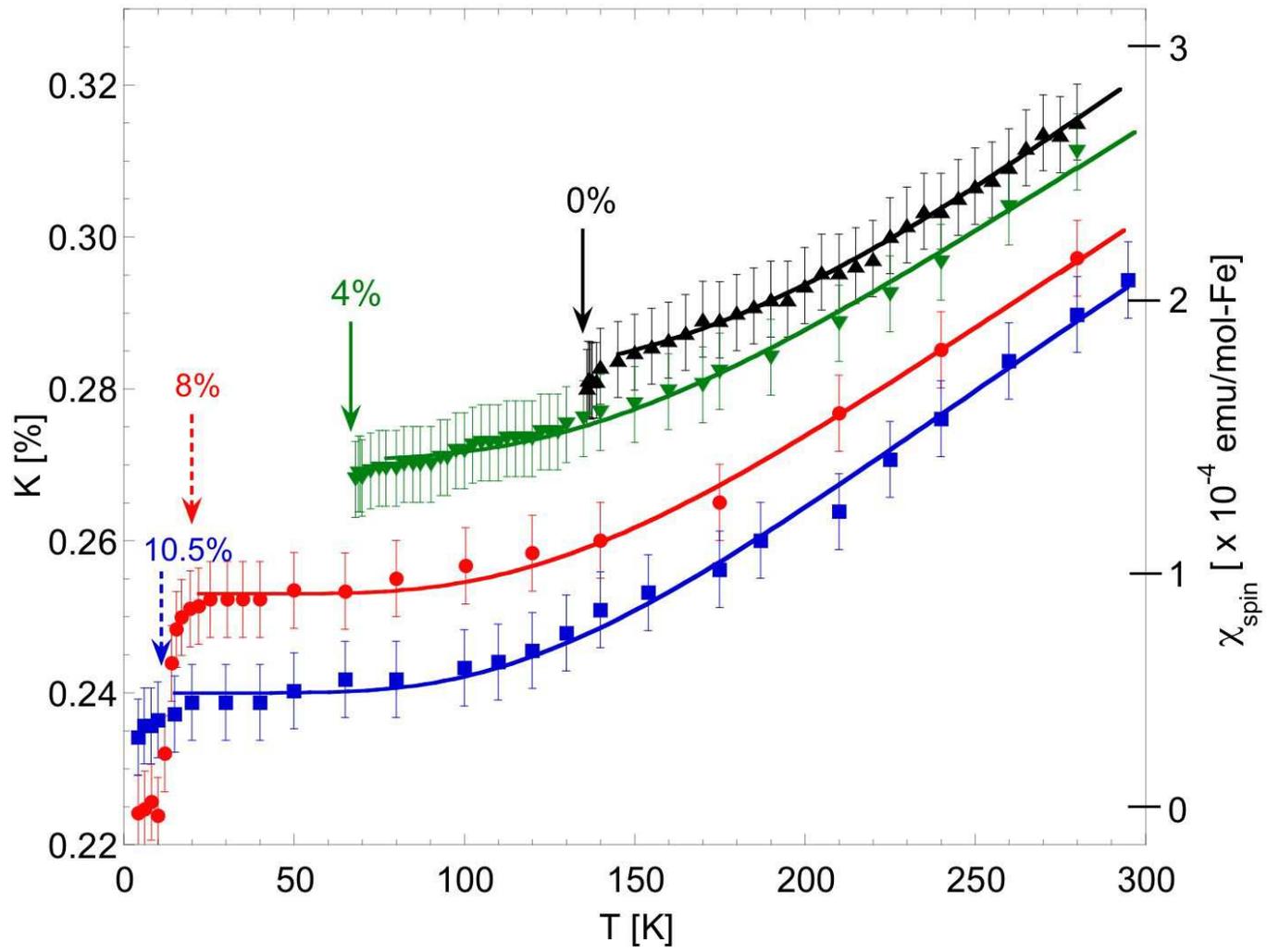
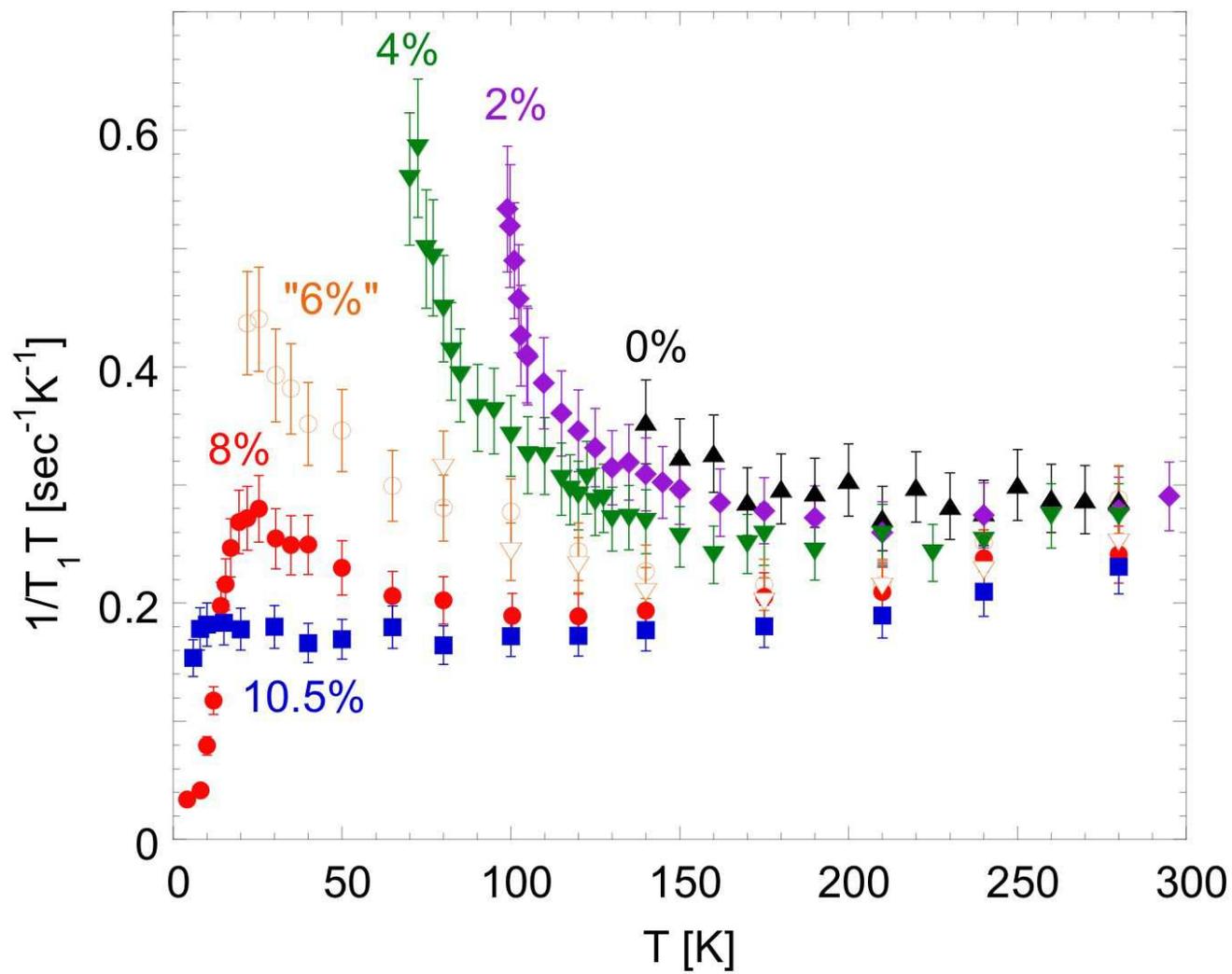
\end{document}